# Searching for effects caused by thunderstorms in midlatitude sporadic E layers

Veronika Barta[a,*], Christos Haldoupis[b], Gabriella Sátori[a], Dalia Buresova[c], Jaroslav Chum[c], Mariusz Pozoga[d], Kitti A. Berényi[a], József Bór[a], Martin Popek[c], Árpád Kis[a], †Pál Bencze[a]

[a] *Research Centre for Astronomy and Earth Sciences, GGI, Hungarian Academy of Sciences, Csatkai str. 6-8., H-9400, Sopron, Hungary*

[b] *Department of Physics, University of Crete, GR-71003, Heraklion, Greece*

[c] *Institute of Atmospheric Physics, Czech Academy of Sciences, Bocni II 1401, 141 31, Prague, Czech Republic*

[d] *Space Research Center, Polish Academy of Sciences, Bartycka 18A, 00-716, Warsaw, Poland*

*\* Corresponding author. E-mail address: bartav@ggki.hu*

**Abstract**

Possible thunderstorm – sporadic E (Es) layer coupling effects are investigated during two measurement periods, one in 2013 and one in 2014. The analysis was based on ionospheric observations obtained from a Digisonde at Pruhonice, the Czech Republic, an ionosonde at Nagycenk, Hungary, and a 3.59 MHz five-point continuous HF Doppler system located in the western part of the Czech Republic. The latter is capable of detecting ionospheric wave-like variations caused by neutral atmospheric waves generated by thunderstorms. The present study searches for possible impacts on Es layers caused by the presence of two active thunderstorms: one passing across the Czech Republic on June 20, 2013 (19:00 – 01:00 LT), and one through Hungary on July 30, 2014 (11:00-01:00 LT). During these two time periods, presence and parameters of Es layer were inferred from ionograms, recorded every minute at Pruhonice and every two minutes at Nagycenk, whereas concurrent lightning activity was monitored by the LINET detection network. In addition, transient luminous events (TLEs) were also observed during both nights from Sopron, Hungary and from Nýdek, the Czech Republic. A noticeable fact was the reduction and disappearance of the ongoing Es layer activity during part of the time in both of the traversing thunderstorms. The analysis indicated that the critical frequency *foEs* dropped below ionosonde detection levels in both cases, possibly because of thunderstorm activity effects. This option, however, needs more case studies in order to be further substantiated.

## 1. Introduction

The coupling of lightning energy into the upper atmosphere and ionosphere has been a topic of enduring interest since the early times of Wilson (1925) who was the first to point out its significance. Research work in this field intensified in the last few decades after the discovery of a series of momentary luminous phenomena, named transient luminous events (TLEs), which occur occasionally above active thunderstorms in the stratosphere and mesosphere (Franz et al, 1990; Pasko et al., 2012 and references therein). The TLEs, which include mostly sprites and elves, as well as a variety of luminous jets emanating from the top of thunderclouds, manifest in a spectacular way the interaction of lightning-induced electrical and electromagnetic energy with the upper atmospheric medium, up to altitudes as high as 90 km. On the other hand, sprites and elves are also known to be closely associated with transient enhancements in the nighttime ionospheric conductivity above ~70 km, which are identified in the amplitude and phase of narrowband VLF signals that propagate in the earth-ionosphere waveguide. These VLF perturbations, which are named early VLF events, have been studied considerably, also in relation with the TLEs (e.g., see Inan et al. (2010); Haldoupis et al., 2013, and more references therein).

With regards to the ionosphere, it is important to stress that the ionospheric perturbations responsible for the early VLF events, are caused by lighting-induced quasi-electrostatic (QE) fields and/or electromagnetic pulses (EMPs) that impact on the D-region ionosphere that is confined during the night between about 75 and 90 km. For altitudes higher than ~90 km, the lightning induced fields are attenuated rapidly because of increases in electrical conductivity which reduces the electric field relaxation time constant, $\varepsilon_0/\sigma$, drastically. This indicates that for altitudes above the D region, that is, in the E-region ionosphere, the effect of lightning-generated fields is expected to be very small, and thus insignificant. This anticipation is endorsed by modeling lightning EMP ionization effects on the ionosphere, which show that strong EMP action is limited, in generating impact ionization, below ~95 km (Marshall, 2012; Gordillo-Vàzquez et al., 2016, Fig. 10).

A question that rises next refers to the existence or not of any lightning effects in the E region ionosphere, and in particular the Es layers which often may dominate E region during nighttime. Es layers are fairly narrow layers of enhanced ionization which locate mostly in the altitude range between 95 and 125 km. With regards to their generation mechanism of windshear theory (e.g., see Whitehead, 1989), the layers are believed to form at specific vertical windshear convergence nodes where long-lived metallic ions can accumulate under the combined action of ion-neutral collisional coupling and geomagnetic Lorentz forcing. The presence of windshear is attributed mainly to the atmospheric tidal winds, whereas shorter scale variations in Es are presumably caused by wind shears associated with gravity waves (e.g., see Haldoupis, 2011).

Efforts made in earlier years did not identify any definitive effects of meteorological origin, including also that of tropospheric lightning on Es layers (e.g. see Whitehead, 1989, and Scotto, 1995). The first paper that came up with such a specific relationship was authored by Davis and Johnston (2005). These authors applied a superposed epoch analysis to statistically identify a significant intensification and altitude descent of midlatitude Es directly above thunderstorms that was attributed to lightning. This study showed that the effects on Es appeared about 6 and 36 hours after lightning, with this delay explained by the propagation time needed for the storm-generated gravity waves (GW) to reach E region and act positively in enhancing the ion-convergence effect of a windshear on Es layer forming. However, we think is hard to explain the observed large time lags of 6 hours, and particularly that of 36 hours, between lightning and Es; to our knowledge, so large time delays in GW propagation from the troposphere to the ionosphere have not yet been reported in the open literature (see also discussion at the end of the paper). In addition, it is difficult to justify why the gravity waves, which are known to be produced by vertical convection in the storm that also produces lightning, act only positively, and not negatively, on the windshear forming process, therefore causing Es intensification only, but not depletion. On the other hand, an alternative mechanism proposed by the same authors, which involves a possible effect of a vertical electric discharge, is unlikely if it can act at Es layer altitudes, especially since there are no lightning-induced discharges seen to occur above about 90-95 km. Even if it worked, however, this process would act instantly and therefore it cannot explain the observed long time lags of 6 and 36 hours of Es initiation effects caused by lightning. In a later paper, Davis and Lo (2008) applied the superposed epoch methodology on a different data set to report that the lightning effect on Es intensification is due to negative cloud-to-ground (-CG) lighting. Although this observation can be of some importance, it is difficult to conceive how GWs, produced by vertical convection inside a storm, relate to –CG lighting polarity discharges.

In an effort to verify the findings of Davis and Johnson (2005), Barta et al., (2013) attempted a similar study, where they applied the superposed epoch methodology of Davis and Johnson (2005) on lightning data observed in the vicinity of the Rome ionosonde which measures routinely Es layer parameters. They found that, contrary to the results of Davis and Johnson, the Es critical frequency *foEs* decreases after the lightning occurrences in a step like fashion, which suggested a long-lasting decrease in Es layer electron densities. In a more recent study, which again followed the methodology of Davis and Johnson, Yu et al., (2015) used lightning and Es measurements from two ionosonde stations, to carry out a detailed analysis, which confirmed only partially the Davis and Johnson (2005) findings on lightning-induced Es intensification. They found some agreement only for one ionosonde station, which was characterized by a time lag in Es intensification of ~34 hours after lightning. This is difficult to be justified in terms of invoking the role of a gravity wave in the process, acting so late after the lightning has taken place.

In summary, the results of the 3 different and independent studies mentioned above suggest a limited agreement, and a fair degree of controversy. This rises more questions on the validity of a possible and specific relation between lightning and Es layers above a thunderstorm. In addition, it suggests that more work is needed in order for a cause and effect relationship between tropospheric lightning and Es layer formation to be identified and established. This need led to the present work effort that is based on individual event-like cases, instead of applying statistical methods, like the superposed epoch analysis that depends on the test period and location.

In this paper, two cases of dynamic mesoscale convective storms generating intensive lightning are considered, which pass through or nearby two ionosonde stations operating in central Europe. In the analysis, observations are used for detailed comparisons with the purpose of identifying a possible connection between lightning and Es. The observations include accurate cloud to ground lightning data and ionosonde Es layer parameters measured from high time resolution ionograms. Note that during both periods, of June 2013 and July 2014, geomagnetic conditions were quiet. The results obtained here include some reasonable facts and useful hints although we note that the present evidence alone is only indicative and thus it cannot be taken as entirely conclusive.

## 2. Description and processing of observations

The present study is based on: (a) lightning measurements obtained with the European lightning detection system (LINET), (b) ionogram recordings from two ionosonde stations (Pruhonice in the Czech Republic and Nagycenk in Hungary), (c) Doppler shift measurements made with a five-point continuous HF radio system operating in the Czech Republic, and (d) optical observations of TLEs made from Sopron, Hungary, and Nýdek, Czech Republic.

The LINET detection system employs accurate GPS timing and time-of-arrival methods to identify the time of occurrence and location of cloud-to-ground (CG) lightning. Moreover, it distinguishes the CG lighting strokes from intra-cloud lightning (IC), whereas it also identifies the lighting polarity, and thus it measures +CG (positive cloud-to-ground) and –CG strokes. The location accuracy is less than about 150 m, while it also measures the CG stroke peak current with an uncertainty of less than 10% (Betz et al., 2009).

Ionospheric electron density profiles are estimated from the ionograms obtained with ionosondes, which transmit sequential short-wave radio pulses in the 1–15 MHz frequency range, vertically towards the ionosphere. The signals are reflected at altitudes where the transmitted frequency equals the ambient electron plasma frequency. Ionospheric layers (E, Es, F1, F2) are characterized by their peak plasma frequencies (*foE, foEs, foF1, foF2*), which correspond to the maximum electron density of each layer. The layer virtual heights (e.g., *h'Es, h'F*) are calculated from the time delays of the sounding waves propagating at the speed of light (Rishbeth and Garriot, 1969).

Ionosondes measure certain Es layer properties rather accurately. The Es intensity is characterized by the critical frequency *foEs*, which can be estimated from the ionogram; it relates to the layer's maximum electron density, $N_e$, through the simple formula: $foEs \sim 8980(N_e)^{1/2}$ in which $foEs$ is in Hz when $N_e$ is expressed in cm$^{-3}$. A typical example of a nighttime ionogram is given in Fig 1, where also the various characteristic ionogram frequencies are shown. In the case of a nighttime Es the difference between the virtual ($h'Es$) and the true height of the layer is small due to the rather low electron densities of the underlying ionosphere. The blanketing frequency (fbEs) corresponds to the lowest radio frequency reflected by the layer and is a measure of the weakest possible ionization in it.

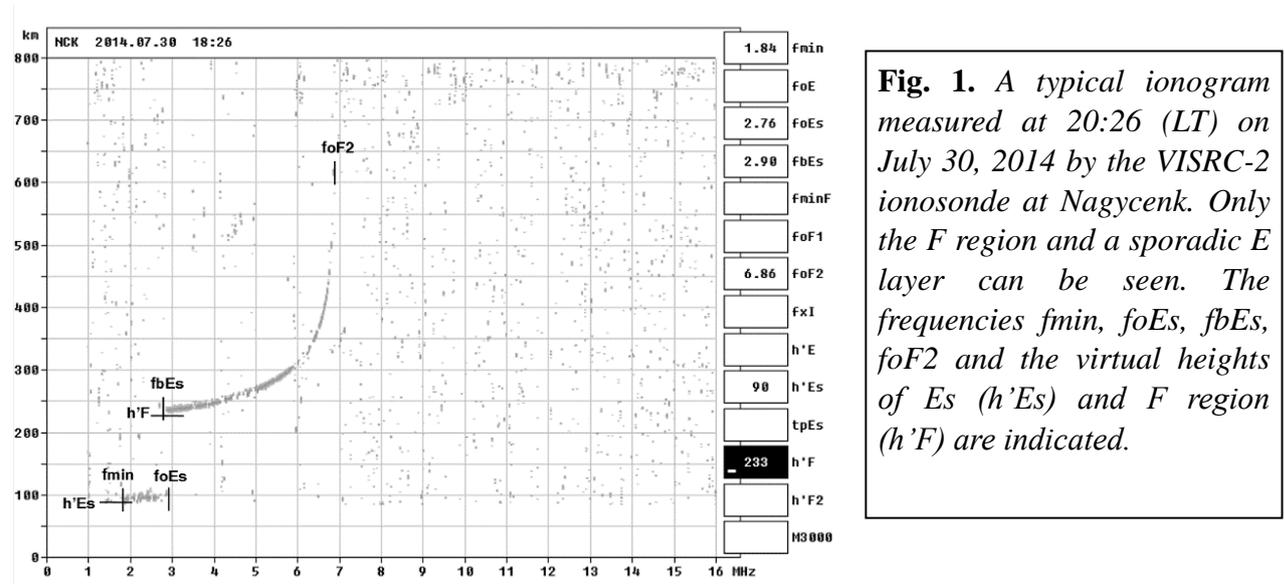

**Fig. 1.** *A typical ionogram measured at 20:26 (LT) on July 30, 2014 by the VISRC-2 ionosonde at Nagycenk. Only the F region and a sporadic E layer can be seen. The frequencies fmin, foEs, fbEs, foF2 and the virtual heights of Es (h'Es) and F region (h'F) are indicated.*

The Es layer measurements used in the present analysis were deduced from ionograms at the two ionosonde stations, in Pruhonice and Nagycenk. The Pruhonice ionosonde, which is located at 50°N, 14.5°E, is a DPS-4D digisonde that operates since 2004, probing routinely the ionosphere every 15 minutes with a 0.05 MHz frequency step. The Nagycenk VISRC-2 digital ionosonde is operated by the Széchenyi István Geophysical Observatory of Hungary, since 2007. In its standard mode, vertical soundings are performed every 15 minutes between 1 and 16 MHz with 0.025 MHz resolution. In the present case studies ionograms were recorded every minute in Prohunice and every 2 minutes in Nagycenk. All the ionograms used here were analyzed manually in order to minimize errors, often caused by automatic scaling.

To investigate possible effects of thunderstorm-produced atmospheric neutral waves on the ionosphere, a multipoint Continuous wave Doppler Sounding System (CDSS) was used. This radio system is operated in western Czech Republic by the Institute of Atmospheric Physics. It is capable of detecting ionospheric fluctuations with periods longer than about 10 s. Such fluctuations may be attributed to infrasound and/or atmospheric gravity waves, both generated in the troposphere by dynamic weather systems. For example, Sindelarova et al. (2009a) used this system to study infrasonic wavelike variations in the ionosphere with period of 2-5 minutes, caused by exceptionally intense convective storms. The operation of CDSS is based on the reflection of the emitted continuous wave signal at a height where its frequency matches the local plasma frequency. In the presence of atmospheric waves acting upon the ionosphere, the plasma is subject to wave motions that Doppler-shift the reflected signal. CDSS is comprised of one receiver and five transmitters spaced over in western part of the Czech Republic, which allows for multipoint spatial measurements to be made. CDSS operates near 3.59 MHz, with the specific frequencies of the individual transmitters being shifted mutually by 4 Hz to display the received signals from all the transmitters in one common Doppler shift spectrogram. The reflection heights of the CDSS signals are estimated by using concurrent DPS-4D ionograms measured at Pruhonice. As shown by Sindelarova et al. (2009b) and Chum et al., (2010, 2014), the estimated phase velocities and propagation directions of AGWs are obtained from the time delays between the observations of the interrelated waves at different reflection points that correspond to the various sounding paths of transmitter-receiver pairs.

Finally, TLEs images (mostly sprites) were captured at Sopron (16.6°E, 47.7°N), Hungary and Nýdek (18.8°E, 49.7°N), the Czech Republic, using low light-sensitive Watec 902H2 Ultimate cameras with F0.8 and F1.3 lens, respectively. A GPS video time inserter provided accurate time stamps for the observation times of the captured events with a 20 ms accuracy. At Nýdek, time was less accurate because the recording system was synchronized manually with a time server on the internet. In this way, the time uncertainty of the video recordings was within a few seconds.

## 3. Observations

### 3.1. Event A, June 20, 2013

A large thunderstorm (supercell) came from southwest in the Czech Republic on the afternoon of June 20, 2013 and crossed through the western part of the country during the late evening and night, as seen on the lightning map in Fig. 2. LINET recordings, located within a range of 200 km from the ionosonde station, were considered and analyzed. It is worth mentioning that in the most active period of the thunderstorm (from 18:00 to 21:00 LT), the number of CG lightning strokes was about 70,000 strokes/hour, that is, about 350 strokes/hour/km$^2$. The temporal variations of the observed parameters are plotted in Fig 3, with panel 3c showing the polarity and peak current of the lightning strokes.

We note that 30 sprite events were captured between 21:17 LT and 23:02 LT from Sopron, Hungary, all caused by +CG discharges. The optical observations started near 21:15 LT and continued during the whole night. The geographic locations of the sprite-causative +CG strokes are indicated in Fig. 2 by the blue stars, while they are signified by red color lines in Fig. 3c; the polarity of all these strokes is positive while their peak currents vary in the range 16.8-186.6 kA.

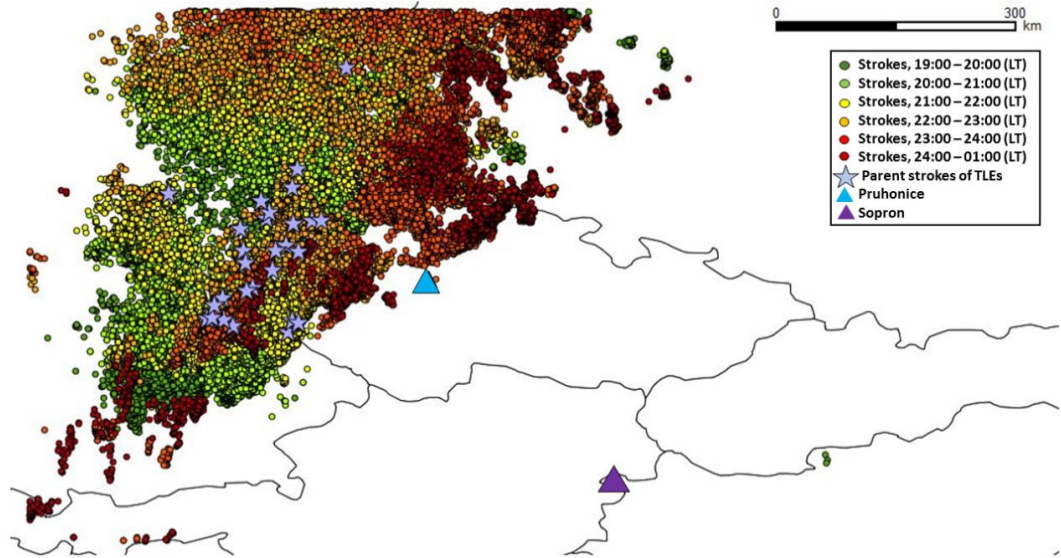

**Fig. 2.** *The geographic locations of the lightning strokes, which occurred near Pruhonice during the observation period from 2013. 06. 20. 19:00 to 2013. 06. 21. 01:00 LT (CET) are represented by dots. The occurrence time of lightning strokes is color coded, indicating that the storm moves towards northwest, over the western Czech Republic. The blue stars show the locations of the sprite-causative +CG discharges. The blue and violet triangles show the locations of the digisonde at Pruhonice, and the sprite camera at Sopron, respectively.*

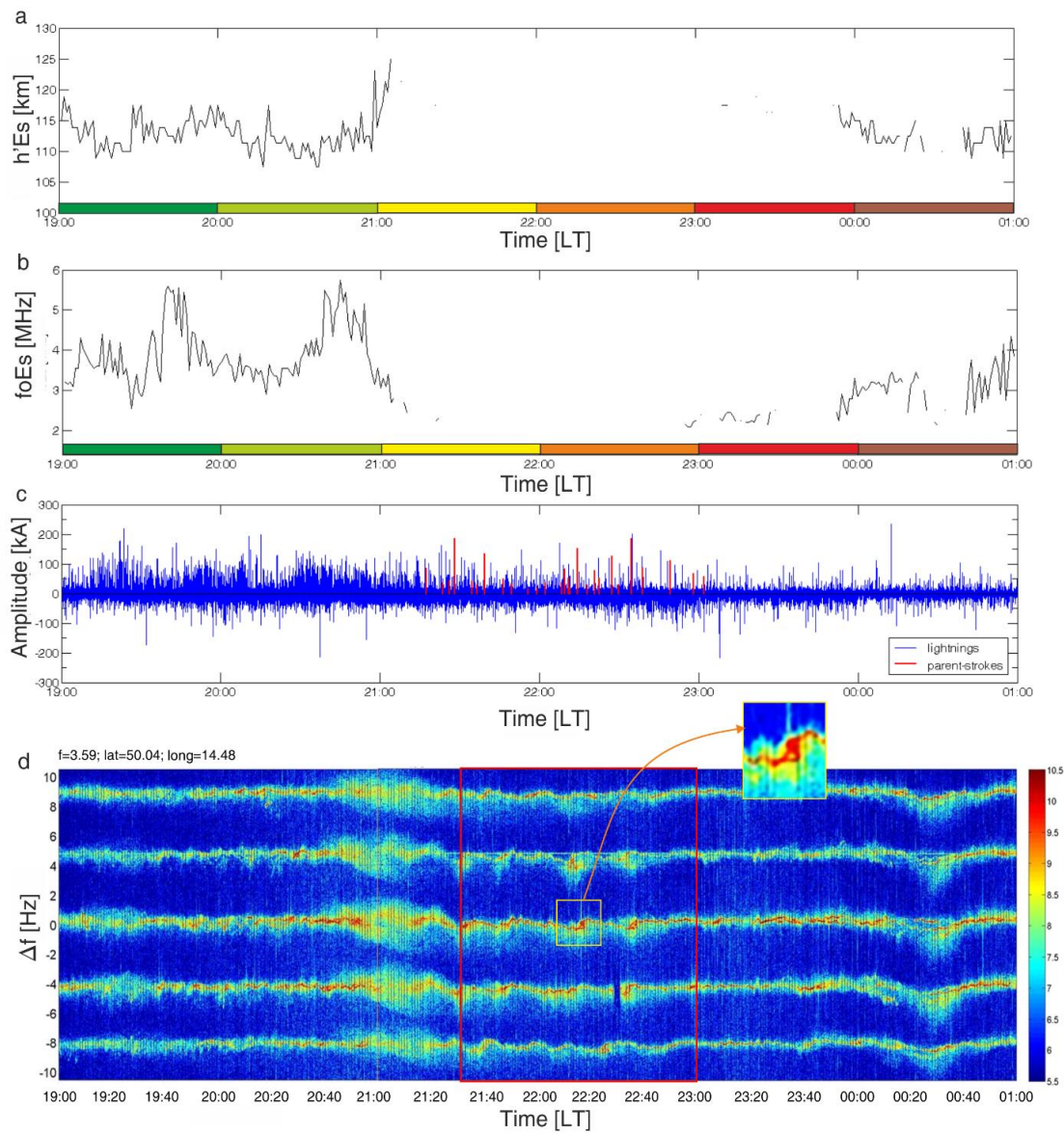

**Fig 3.** *Temporal variations of (a) h'Es; (b) foEs; (c) peak current and polarity of CG strokes within 200 km from Pruhonice with the sprite-causative discharges shown in red; (d) the CDSS Doppler spectrogram. The colored time intervals at the bottom of the upper two panels correspond to the color coded times of lightning discharges in Fig. 2. Doppler shifted signals in the time interval 21:30-23:00 LT, inside the red rectangular form S-like traces (as it is magnified in the inset), which are attributed to a wavelike modulation caused by gravity waves (Chum et al., 2010)*

Figure 3a and 3b show the Es layer virtual height and critical frequency, *h'Es* and *foEs*, which refer to the layer's height, and strength or intensity, respectively. The color coded bars at the bottom of each plot in the upper two panels correspond to the colored positions of the lightning strokes shown in Fig. 2, thus one may estimate their distance from the ionosonde during consecutive times of observation. Inspection of Fig. 3 shows that the Es layer activity disappeared for a couple of hours, mostly during the time when the sprites occurred. The lack of Es activity indicates that the layers became weak; having their electron density dropped below ionosonde detection levels (~1.2x10$^4$ cm$^{-3}$ which corresponds to the ionosonde 1 MHz frequency threshold).

As mentioned in the introduction, a thunderstorm may affect the upper mesosphere and lower thermosphere, and therefore the Es layer formation in particular, through the action of storm-generated gravity waves. It is worth pointing out that, during event A, the CDSS Doppler sounder in western Czech Republic detected gravity wave like variations. This was implied by the Doppler spectrograms shown in the bottom panel of Fig. 4. The observed wave-like variations in Doppler shift in the time interval from 21:30 to 22:50 are indicative of short period GWs, while those between 21:40 and 22:10 are possibly due to infrasound waves. This suggests that the source of the waves could be relatively local, occurring in a likely relationship with the passage of the storm. The Doppler system is installed in the west part of Czech Republic where the thunderstorm passed through from ~ 20:00 to ~ 23:00, as seen from the storm time history in Fig. 2. The time interval 21:30-23:00 (LT), inside the red rectangular box in Fig. 3d, was selected for analysis, because clear traces of gravity wave modulation in the form of S-shape (Chum et al. 2010) were present in the Doppler spectrograms. It is interesting to point out that this GW signature coincides with the disappearance, or drastic weakening, of Es, thus the CDSS signal at 3.59 MHz during this time was likely passing through the E region to be reflected in the F layer. The analysis showed that GWs propagated in the North-East direction (azimuth ~45°), which endorses the possibility that their source is the thunderstorm that came from southwest to transverse through the western part of the Czech Republic.

**3.2 Event B, July 30, 2014**

Event B refers to a storm that occupied a large area, moving across Austria, the west part of Hungary and part of Slovakia, during most of July 30, 2014 (Fig. 4). LINET data on CG lightning occurring within a circle of 200 km radius centered at the ionosonde were considered in the analysis. The geographic locations of the lightning strokes which occurred in consecutive color-coded hourly intervals are shown in Fig. 4. The temporal variations of CG stroke polarities and peak currents are shown in Fig. 5c. The number of lightning strokes exceeded 45,000 strokes/hour (225 strokes/hour/km$^2$) during the most active period of the thunderstorm, from 15:00 to 16:00 LT.

During the time interval from 20:39-00:07 (LT), 25 sprites were captured from the camera at Nýdek, in the Czech Republic. The sprite-causative discharges, and thus their polarity, were not identified in this case because of the time inaccuracies in the Nýdek measurements. Nevertheless, the location of the sprites was estimated here from the direction and distance of the events from the observing site. The direction of the events was determined by matching the stars in the camera field of view with the star map (for details see Mlynarczyk et al., 2015). The distance of the events was estimated from the elevation angles of sprite elements in the video by taking into account that sprites are confined in height mostly between 50 and 90 km (Pasko et al., 2012). In this way, the estimated geographic locations of the observed sprites are depicted by the blue stars in Fig. 4, while the approximate occurrence times shown with red arrows in Fig. 5c.

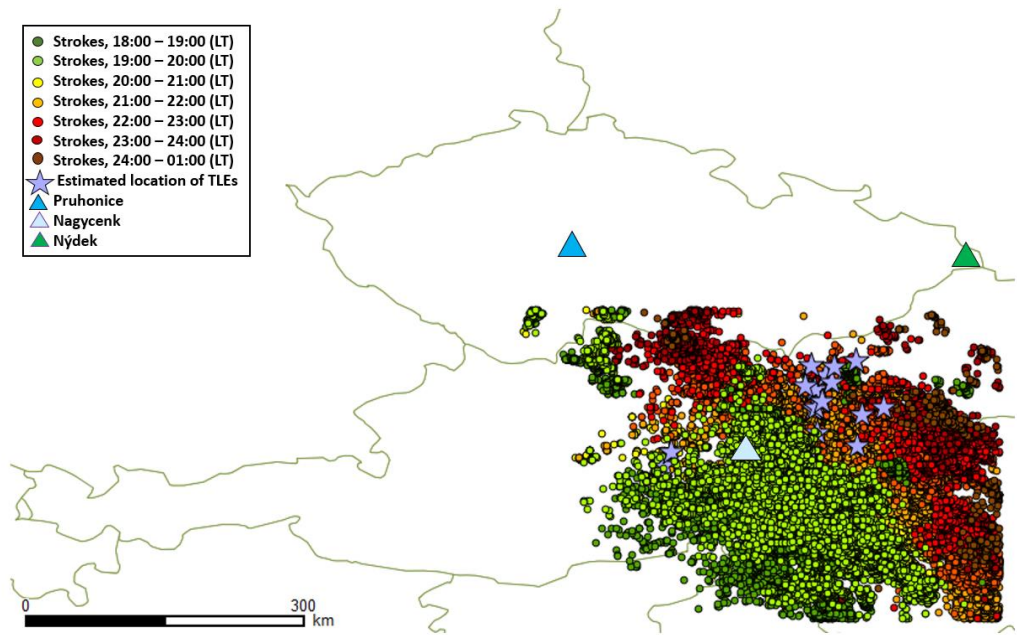

**Fig. 4.** *Geographic locations of lightning strokes during event B from 2014. 07. 30. 18:00 to 2014. 07. 31. 01:00, LT. Sequential time intervals are color coded. The estimated locations of the observed sprites are shown with blue stars. Ionosonde stations at Pruhonice and at Nagycenk are indicated by the blue and light-blue triangles, respectively. The green triangle shows the location of the sprite camera in Nýdek.*

The time series of *h'Es* and *foEs* are summarized in Fig. 5a and 5b. The ionograms used here for computing the Es parameters were recorded by the ionosonde at Nagycenk (its location indicated by the light-blue triangle in Fig. 4) every two minutes from 11:00 to 01:00 LT. Doppler spectrogram data was not considered in this case because the thunderstorm occurred relatively far (~250 km) from the area monitored by CDSS. . Similarly as in the case of the event A, the Es layer also disappeared in event B, apparently because its electron density dropped below the ionosonde detection level.

In both events A and B there were no characteristic changes in the following lightning discharge parameters: number frequency (strokes/minute), average peak current, CG/IC ratio, and +/- polarity ratio of lightning during the period of the disappearance of Es, relative to the values occurring before and after the loss of Es.

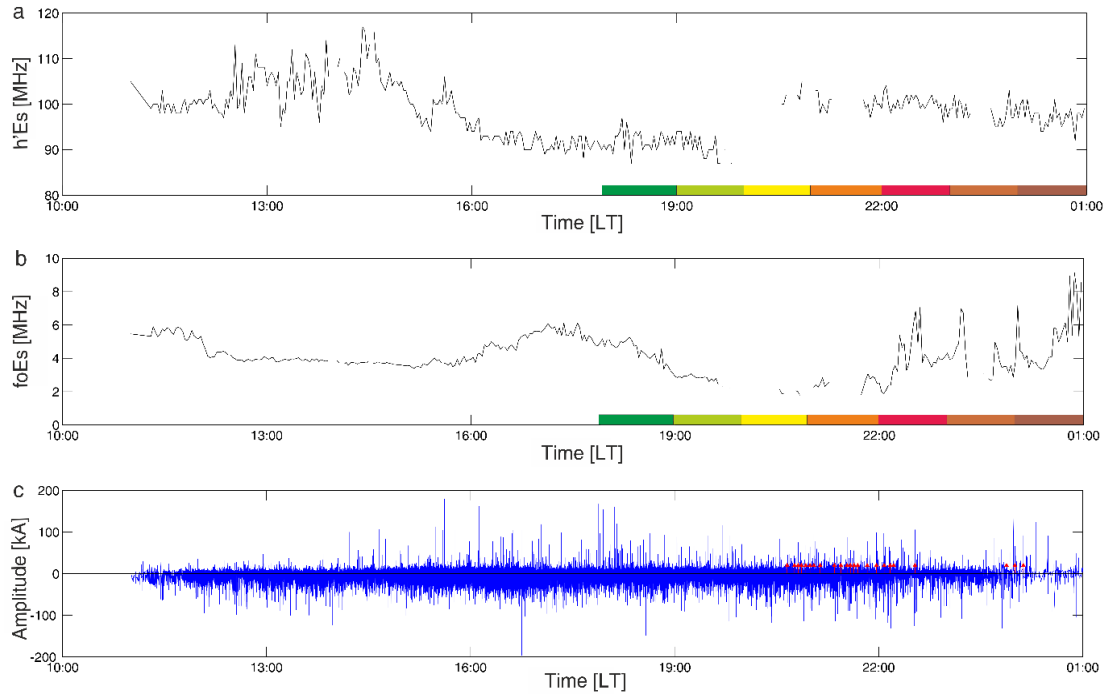

**Fig. 5.** *Temporal variations of: (a) h'Es, and (b) foEs observed by the Nagycenk ionosonde on July 30, 2014 (colored time intervals at the bottom of each plot correspond to the sequential color coded times shown in Fig. 4), (c) peak current and polarity of CG strokes within 200 km from Nagycenk, with the red arrows showing the times of sprite occurrences.*

### 4. Discussion

In the following 3 subsections, the word "effect" refers to the observed disappearance of Es during at least part of the time the thunderstorm passed through or nearby the ionosonde location.

**4.1 Is it a regular or irregular effect?**

In order to decide whether the Es disappearance during the thunderstorms represents an irregular situation and not a regular diurnal characteristic, the variability of *foEs* was analyzed and inspected for a few days before and after June 20 (event A), when LINET did not measure any lightning activity in the vicinity of the ionosonde location. The *foEs* variations observed at Pruhonice from June 18 to 23 are plotted in Fig. 6. As seen, it is apparent, that during this whole 5 day period, the Es layer was detected always except for the time when the thunderstorm occurred in June 20, as discussed above and shown in Fig. 3. The same behavior, as in event A, was also observed by the Nagycenk ionosonde for a few days before and after event B (not plotted here). This is, the substantial electron density reduction in Es and its disappearance during the thunderstorm in event B does not occur during the same time interval in the storm-free days. These comparisons hint the existence of a possible relationship between a thunderstorm and the weakening of Es, although this evidence alone is far from being conclusive.

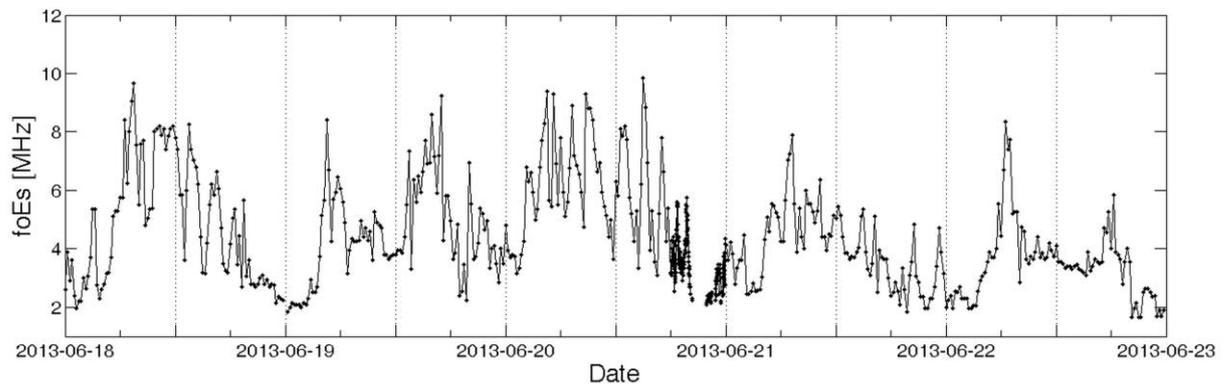

**Fig. 6.** *Variation of foEs observed at Pruhonice during 5 days, form June 18 to 23, that is, two days before and after the day of June 20 when event A was observed in the presence of a thunderstorm.*

**4.2 Local or regional effect?**

In order to examine further the local nature of the phenomenon, *foEs* and *h'Es* time series, obtained from other, far away, ionosondes in Europe were collected and analyzed for the campaign periods and compared with the results showed in Fig. 2 and 4. The purpose of this analysis was to decide whether the observed depletion in Es is due to effects in the ionosphere of local geographic origin, for example in relation with the passing thunderstorm, or due to some other, larger scale regional effects. The observations from 4 additional European ionosondes are compared in Fig. 7 with the Pruhonice and Nagycenk Es layer measurements during event A. Despite the large differences in Es variation among the stations, *foEs* was relatively low but present in the records of most European ionosonde stations during the evening of June 20. This contrasts with the observations at Pruhonice where the digisonde did not detect any Es traces for part of the time when the thunderstorm was present in its vicinity. The results of analysis for event B are similar as those of event A, this is, Es was undetected only at Nagycenk for at least part of the time when the thunderstorm passed through the ionosonde.

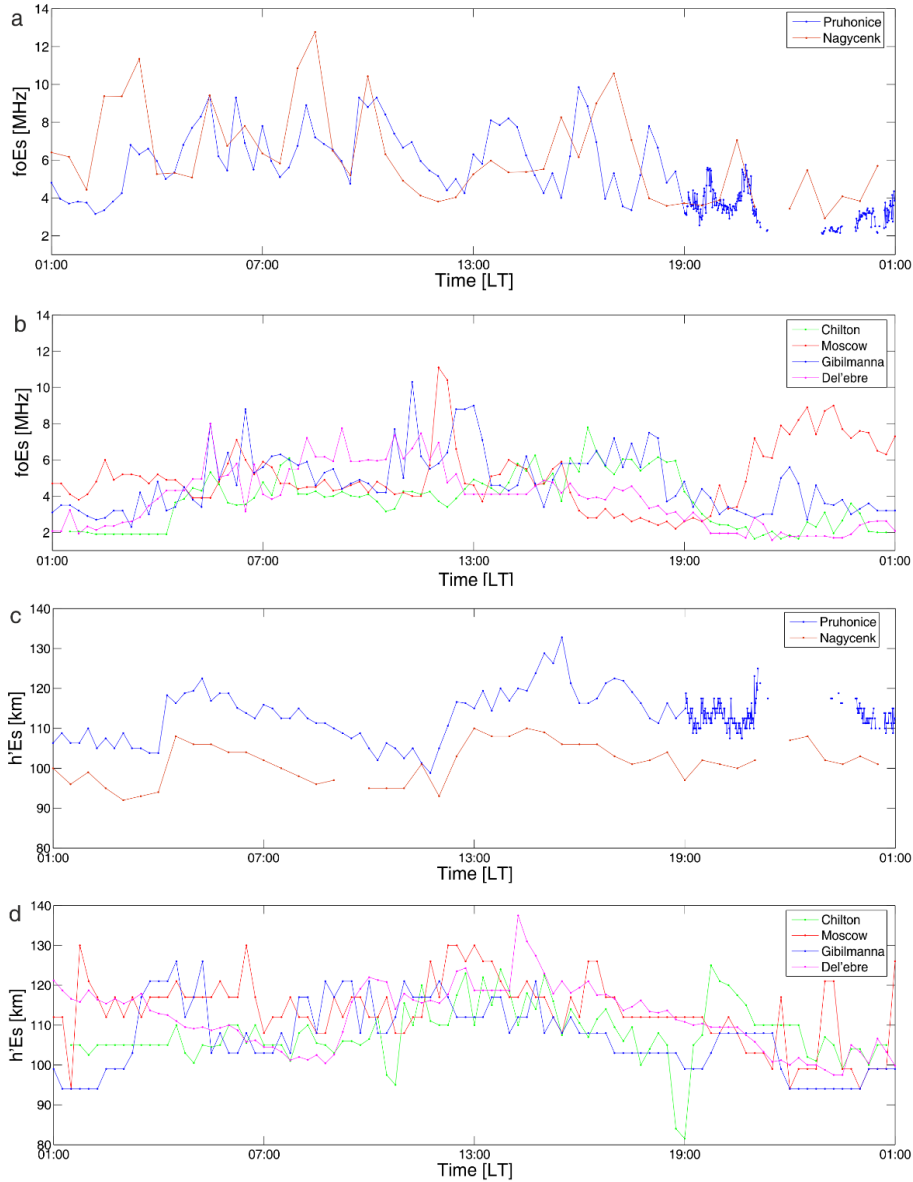

**Fig. 7.** *The variation of (a) foEs and (c) h'Es at Pruhonice and Nagycenk compared with recordings of various other European ionosondes (b and d) during event A of June 20, 2013.*

### 4.3 Is the effect caused by Es layer descent?

A decrease of electron density inside Es is known to accompany the altitudinal descents of the layers, which occur regularly. The Es descent is followed by a decrease in metal ion density because of increases in recombination reaction rates with decreasing altitude. This is because recombination requires three body collisions which are more likely to happen at lower altitudes than at higher (Haldoupis, 2012). As shown in Fig. 7c, the variations in layer height *h'Es* at Pruhonice are similar, with those at Nagycenk, however, being systematically lower than at Pruhonice. This might be an instrumental effect, because otherwise it is difficult to be explained, given the nearness of the two stations.

As seen, *h'Es* at Pruhonice was lower in the first part of the day than in the measurement campaign period during evening and nighttime hours. On the other hand, *h'Es* increased before the disappearance of Es, between about 20:00-20:30 LT. This suggests that the observed reduction in Es electron density cannot be due to layer descent, as discussed above, and thus its relationship with a regionally local effect, possibly the ongoing thunderstorm, is plausible.

**4.4 Possible physical mechanisms**

In summary, the substantial electron density reduction in *foEs* that was detected during the passage of the thunderstorms appears not to be there during the same times in the storm-free days before and after the present events under consideration. Furthermore this reduction appeared to have been caused by a regionally local effect as it cannot be explained by layer descent in altitude. These evidences point to the possibility that the reduction in Es electron density is due to effects caused by the passage of the nearby thunderstorm.

By adopting the option of a relationship between the thunderstorm and Es, one needs to provide a physical explanation of why and how the Es electron density is affected in relation with the nearby thunderstorm activity. In our present understanding, the interaction of thunderstorm energy with the mesosphere and thermosphere is attributed to two main independent processes which initiate in the thunderstorm and subsequently impact onto the upper atmosphere and ionosphere. These include: (a) the lightning-induced quasi-electrostatic (QE) and/or electromagnetic pulsed (EMP) fields which act on the upper atmosphere-ionosphere instantly, and (b) the neutral atmospheric waves, mostly gravity waves (GW) which are generated by free convection inside the storm and have their energy moving upwards, therefore affecting, after a certain propagation time, the ionospheric plasma directly through collisional coupling between the ions and neutrals.

We comment first on the possibility of QE and/or EMP effects on sporadic E layers. The detection of TLEs (e.g., mostly sprites) in both events A and B indicate the presence of rather large QE fields, and possibly EMP fields as well, in the mesosphere and lower thermosphere above the storm; also they associate to large charge moment changes (CMC) associated with their causative lightning discharges (e.g., see Huang et al., 1999; Hu et al., 2002; Satori et al., 2013). Can such large CMCs and their corresponding QE fields, which are capable at times of producing TLEs in the upper D region, have an effect on E region where Es is situated? As discussed in the introduction, the mechanisms that involve both QE and EMP, fields, are very unlikely to have direct effects on Es. This is because Es resides at altitudes mostly between 100 and 120 km, where the lightning-produced fields are reduced greatly because of the diminishing effect on them of the enhanced electrical conductivity above D region heights, that is, above 90 to 95 km (e.g., see Gordillo-Vàzquez et al., 2016).

On the other hand, one cannot exclude other indirect processes of EM coupling for explaining the observed disappearance of Es during the thunderstorms. For example, VLF observations have shown that during strong lightning conditions there exist long lasting elevations in D region electron densities, caused by electron impact of EMP fields emitted from large peak currents of CG lightning discharges (Haldoupis et al.

2012, 2013). If this situation is present above an active thunderstorm for long time, then is possible that it leads to enhanced absorption of the ionosonde HF pulses, which then may not reach Es reflection heights, therefore an Es layer in this case can become undetected. In order for this idea to be a viable option, however, more studies are needed which involve in addition to the present measurements narrow band, subionospheric, VLF measurements.

If thunderstorm QE and EMP coupling effects on Es are to be excluded, the second impact caused by an interaction of thunderstorm-produced AGWs with Es layer can be considered. The most likely reason for this interaction to take place relies on the windshear theory of Es generation and the positive or negative role that GWs may play on layer formation through their vertical wind shears associated with their horizontal zonal and meridional wind components. This GW role is widely recognized and accepted in the long going Es research (e.g., see Whitehead, 1989, Mathews, 1998, and Haldoupis, 2011). More recent research on Es layers, by means of using the incoherent scatter radar at Arecibo and improved ionosonde techniques, established that both Es formation and altitudinal layer descent with time are controlled mostly by wind shears provided by solar tidal winds, particularly the diurnal and semidiurnal tides (Mathews, 1998; Haldoupis, 2012). On the other hand the effect of gravity waves on Es formation and descent (which affects both *foEs* and *h'Es*) may cause shorter scale Es variability with time scales less than a few hours. In this case, GWs can act positively (increasing) or negatively (decreasing) the prevailing tidal plasma forcing and thus can influence tidal Es layers. As discussed by Haldoupis (2012), strong amplitude GWs can intervene and alter the regular tidal Es-forming process in reinforcing or disrupting its course, a phenomenon that is often seen in ionogram recordings, and categorized as spread Es.

In the framework of the thunderstorm – Es coupling mechanism under consideration, the observed Es disappearances in both events A and B, may represent examples of Es disruption caused by GWs, presumably generated by the passing thunderstorms. However, if this mechanism is valid then one should expect to observe also cases when the thunderstorm GW activity would act positively to reinforce ongoing Es layer activity, that is, to increase the *foEs* critical frequencies and thus enhance the layer plasma densities. This was not observed in the present data, thus it remains a challenge that needs to be met. The authors, therefore, suggest of making further experimental studies of the type carried out in the present work to possibly gain more insight into the role of thunderstorm-generated GWs impacting onto Es layers.

The potentiality of the GW mechanism is endorsed by the CDSS Doppler spectrograms used in event A (storm of June 20, 2013). These indicated the presence of gravity wave-like ionospheric variations, whose direction of propagation was in good agreement with the directional motion of the storm. Another fact in favor of the assumed thunderstorm GW - Es coupling mechanism, which seems to be supported by the present data, is that the storm does not necessarily need to pass exactly through the ionosonde monitoring area, because the storm-generated GWs propagate both vertically and horizontally. Along this line of reasoning, the GWs reaching the lower thermosphere can excite secondary neutral atmospheric waves which can be trapped in the lower thermosphere and propagate over large distances (Vadas et al. 2003;

Snively and Pasko 2003; Sindelarova et al. 2009b, Snively et al., 2010; Nishioka et al., 2013). One more point worth mentioning, which appears to complicate the effectiveness of the GW - Es interaction mechanism, is the well known filtering effect on GWs that may be caused by neutral winds in the upper mesosphere, which in turn can reduce GW energy penetration into the E region (e.g, see Cowling et al., 1971; Vadas, 2007). Since in the present study we have no direct wind measurements in the upper atmosphere, we cannot speculate further on this possibility.

Finally, we comment briefly on the times required for the thunderstorm-generated GWs to reach E region and impact on Es plasma. The evidence here showed that in both events A and B, there is 1 to 2 hours delay of the Es disappearance relative to the appearance of the storm near the area covered by the ionosonde station. As discussed in the following paragraph, this possible delay agrees with the results of several studies indicating that storm-generated atmospheric waves reach E region heights in times ranging from about 20 min to two hours. On the other hand, the existence of such a time delay argues strongly against the supposition that thunderstorm-induced QE and EMP fields affect Es, because if this were to be taking place it had to occur instantly.

The GW propagation delays mentioned above, have been confirmed in several studies. For example: Suzuki et al (2007) used an all-sky airglow imager to find that the propagation of storm generated GWs from 10 km to 96 km height, that is, at OH emission altitudes, took ~48 minutes. On the other hand, Xu et al. (2015) detected AGW undulation activity by all-sky airglow imaging at 87 km height, about 40 minutes after a severe meteorological convective cell passed nearby. Also, Azeem et al. (2015) tracked AGWs that originated in a convective storm in the troposphere to the stratosphere, mesosphere, and ionosphere. They observed concentric wave-like forms in the nightglow layer near 85–90 km within only 15 min after a storm passed underneath. On the other hand according to simulations by Snively and Pasko (2003), the GWs generated at the troposphere can reach Es heights (100-120 km) in 1.0 – 1.5 hours later. In summary, the estimated GW propagation times from the troposphere to the E region range from about 20 minutes to 2 hours, which is in fair agreement with the present observations. It is also worth mentioning that such propagation times are much more reasonable than the times of 6 and 36 hours found in the superposed epoch analysis of Davis and Johnson (2005) and Yu et al. (2015).

Finally, the present studies show that the observed Es disappearances occurred about the same time interval when several sprites did also occur. Based on the above discussion and the adoption of the GW - Es coupling mechanism at work, there is no obvious physical reason for this to be happening. One possible connection however, can be that the strong inhomogeneities in the mesosphere required to trigger sprite discharges as postulated by Pasko et al., 1997, and Fadnavis et al. 2009, among several others are caused by the presence of storm-generated gravity waves and gravity wave breaking in the E region heights.

## 5. Summary and conclusions

Thunderstorm-ionosphere coupling was investigated using two case studies in which thunderstorms moved near two ionosonde stations in Central-Europe on June 20, 2013 and on July 30, 2014. During both cases the ongoing Es layer activity, monitored by the ionosondes, ceased for part of the time when the thunderstorms passed by. The results implied that the Es layer disappearance relates to a weakening of the electron density of the layer to below about $1.2 \times 10^4$ cm$^{-3}$, which corresponds to plasma frequencies lower than the ionosonde frequency threshold of 1 MHz. Further evidence indicated that the disappearance of Es was observed only by the stations near the storm, and not elsewhere in far away stations. This suggests the possibility of a cause-and-effect relation between a thunderstorm and an overlying Es layer. Given this option, we note that the Es disappearances initiated about 1 to 2 hours after the storms approached the area monitored by the ionosondes agree with anticipated propagation times of GW energy from the troposphere to the E region ionosphere.

In conclusion, the present results postulate the possible existence of a cause-and-effect relationship between active thunderstorms in the troposphere and overlying Es layers in the lower E region ionosphere. It was suggested that the mechanism responsible for this interaction relies on the action of gravity waves initiated by convective motions in the thunderstorm. These GWs can propagate in the mesosphere and lower thermosphere and into the E region where they may act to disrupt Es layer formation. This can be done through the GW vertical wind shears which act to reduce plasma accumulation into a layer, in accord with the principles of windshear theory that applies in Es layer generation.

Finally, it is important to stress that the results here are tentative because the present evidence is only indicative and therefore cannot be taken as conclusive. More studies of the type described in this work are needed in order for the hinted relation and interaction mechanism between thunderstorms and Es layers to be firmly established.

**Acknowledgments**

The communication between the authors was facilitated by the scientific program TEA-IS of European Science Foundation. The lightning data were kindly provided by LINET (Nowcast GmbH). D. Buresova and J. Chum acknowledge support under the grants 15-07281J and P209/12/2440 by the Czech National Foundation. This work was also supported by the National Research, Development and Innovation Office, Hungary-NKFIH, K115836 and by the TAMOP-4.2.2.C–11/1/KONV-2012-0015 (Earth-system) project sponsored by the EU and European Social Foundation furthermore by the TAMOP 4.2.4. A/2-11-1-2012-0001 'National Excellence Program'. Contribution of J. Bór was supported by the János Bolyai Research Scholarship of the HAS. M. Popek acknowledges support from the grant GACR 14-31899S. Additionally, the authors wish to thank Eric Kihn and the Space Physics Interactive Data Resource at the National Geophysical Data Center for providing Web-accessible *foEs* measurements from European Digisonde stations.